\documentclass[3p,times]{elsarticle}

\usepackage{amsmath,amssymb,amsfonts}

\usepackage{graphicx}

\newcommand{\be}{\begin{equation}}
\newcommand{\ee}{\end{equation}}
\newcommand{\ba}{\begin{eqnarray}}
\newcommand{\ea}{\end{eqnarray}}
\newcommand{\ban}{\begin{eqnarray*}}
\newcommand{\ean}{\end{eqnarray*}}

\begin{document}

\begin{frontmatter}

\title{Parton Energy Loss in Strongly Coupled AdS/CFT}

\author{Berndt M\"uller}

\address{Department of Physics, Duke University, Durham, NC 27705}

\begin{abstract}
This is a brief review of the theory and phenomenology of parton energy loss in strongly coupled field theories with a gravity dual and its comparison with parton energy loss in QCD at weak coupling.
\end{abstract}


\end{frontmatter}


\section{Introduction}
\label{sec:intro}

The use of jet related cross sections as a probe of the hot QCD matter produced in relativistic heavy ion collisions rests on the fundamental assumption that the hard scattering process can be factorized from the final-state interactions of the scattered hard parton and its fragmentation into hadrons. In the collinear factorization formalism the double differential cross section for the hadron spectrum takes the schematic form
\be
\frac{d\sigma_{h}}{dy\, dp_T^2} \sim \sum_{a,b,c} \int f_1(x_a) f_2(x_b) \frac{\hat\sigma_{ab\to c}}{d\hat{t}} \tilde{D}_{c\to h}(z) ,
\label{eq:jetX}
\ee
where the medium modified fragmentation function $\tilde{D}_{c\to h}(z)$ encodes the interaction of the outgoing parton with the medium and the hadronization process. I will assume that a formula of the form (\ref{eq:jetX}) holds  whether or not the QCD medium is a strongly or weakly coupled system. (See additional comments below concerning the validity of this assumption in a strongly coupled gauge theory.) 

The perturbative QCD (pQCD) approach to jet quenching further assumes that the evolution of the jet itself prior to hadronization is governed by perturbative dynamics and can be factorized from the medium. In contrast, it does not matter whether the dynamics of the medium is governed by strong or weak coupling; its effect on the jet evolution can be encoded in various transport coefficients that can be expressed as matrix elements of certain operators, which may or may not be calculated perturbatively. The two most important transport coefficients relevant to jet quenching are \cite{Majumder:2008zg}:
\be
\hat{q} = \frac{\langle p_T^2 \rangle_L}{L}, 
\qquad
\hat{e} = \frac{\langle \Delta E \rangle_L}{L}.
\ee
$\hat{q}$ expresses the rate of transverse momentum broadening per unit path length of a hard parton and controls the radiative energy loss; $\hat{e}$ denotes the rate of longitudinal energy loss by absorption of virtual gluons from the medium (elastic collisions). In the common pQCD formulation radiative and elastic energy loss are treated separately. The general pQCD theory of radiative energy loss, due to Zakharov \cite{Zakharov:1996fv} and Baier {\em et al.} \cite{Baier:1996kr} (BDMPS), was discussed by S.~Caron-Huot \cite{CaronHuot:2010HP}; an overview of its various implementations was given by van Leeuwen \cite{Leeuwen:2010HP}. Two important phenomenological features are the suppression of the medium contribution to radiation at early times due to interference with radiation from the initial hard scattering process and its suppression at late times due to coherence effects (LPM effect \cite{Landau:1953um}).

\begin{figure}
\label{fig:comp}
\centerline{\includegraphics[width=0.6\textwidth]{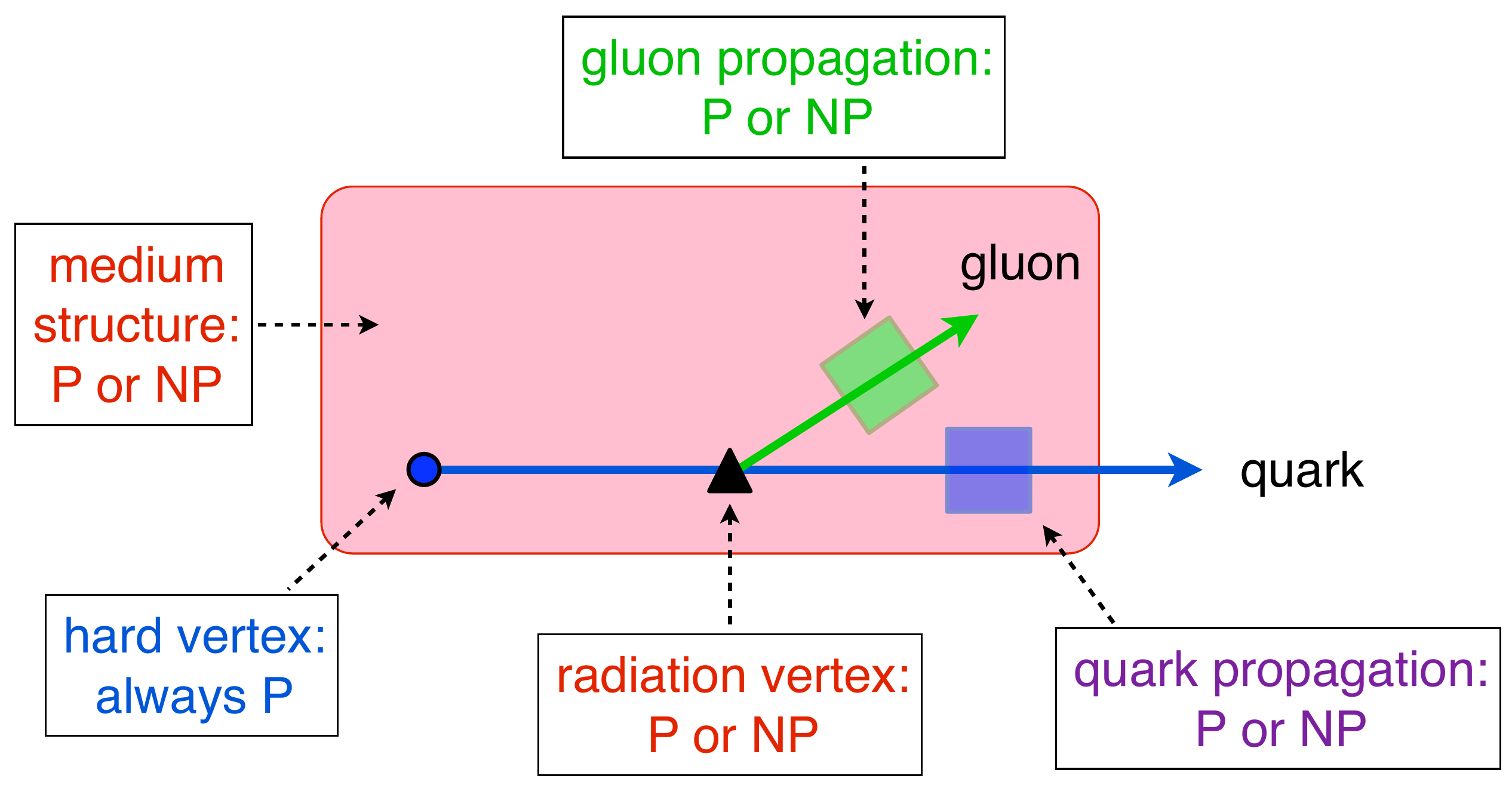}}
\caption{Schematic diagram showing the various components of jet quenching in hot QCD matter. The hard scattering vertex always must be treated as P. The pQCD approach treats all components as P, with the exception of the factorized structure of the medium, which can be NP. In a strong coupling scenario of jet quenching also the radiation vertex, the propagation of the hard parton (here taken as a quark), and the dynamics of the radiated gluons will be NP. [P = perturbative; NP = nonperturbative.]}
\end{figure}

When we relax the assumptions of the pQCD jet quenching theory and also consider strong coupling effects between the medium and the jet, three different new aspects come to mind (see Fig.~\ref{fig:comp}): ($i$) the radiation vertex, usually encoded in the parton splitting function, ($ii$) the propagator of the hard parton, which may or may not have the eikonal form assumed in pQCD, and ($iii$) the dynamics of the radiated gluons. In a rigorous strong coupling theory, such as the ${\cal N}=4$ supersymmetric Yang-Mills theory (SYM) at strong 't Hooft coupling, all these components of the jet quenching process are treated nonperturbatively. We now turn to the discussion of jet quenching in this exactly solvable theory. 

\section{Jet quenching in AdS/CFT at strong coupling}
\label{sec:AdSCFT}

Most efforts to explore the dynamics of jet quenching at strong coupling make use of the holographic duality between a certain string theory on the 10-dimensional AdS$_5\times$ S$_5$ and the ${\cal N}=4$ SYM theory in 4-dimensional Minkowski space \cite{Maldacena:1997re,Aharony:1999ti}, usually simply referred to as AdS/CFT duality. In the large-$N_c$ limit, the string theory degenerates to a classical supergravity theory on the AdS$_5$ space, which has the Minkowski space as its (infinite) boundary. The large-$N_c$ limit has to be taken in such a way that the elementary gauge coupling $g$ remains small, but the 't Hooft coupling $\lambda = g^2N_c \to \infty$. \footnote{Note that for QCD at a scale of approximately 1 GeV, we have $g\approx 2, N_c=3$ and thus $\lambda\approx 12$. Whether this value is large enough to consider QCD strongly coupled is not known, but it was recently shown for certain operators in the SYM theory that the transition between weak and strong coupling occurs at $\lambda\approx\pi^2$, corresponding to $\alpha_s \approx 0.25$ \cite{Benna:2006nd}.} The AdS/CFT duality maps the quantum dynamics of the SYM on the 4-dimensional ``Minkowski boundary'' onto the classical dynamics of (super-)gravity in 5 dimensional ``bulk''. Somewhat loosely speaking, the radial $5^{th}$ coordinate, here denoted as $\chi$, near the boundary corresponds to the ultraviolet limit of virtuality in the quantum field they, while the deep bulk encodes the quantum dynamics in the infrared (IR).\footnote{Other notations for the radial coordinate commonly used in the AdS/CFT literature are $\chi = z = R^2/r$. Here we follow the notation of Hatta {et al.} \cite{Hatta:2007cs,Hatta:2008tx}.} An easily readable overview of the AdS/CFT duality and its use in the exploration of strong coupling physics can be found in \cite{Gubser:2009md}.\footnote{Other talks at this conference, which explored various aspects of using the AdS/CFT duality to investigate strong coupling effects relevant to hot QCD matter, were given by Y.~Kovchegov, E.~Iancu, and J.~Casalderrey-Solana.}

The supergravity theory on the AdS$_5$ space is dual to the SYM theory in the vacuum state. The thermal gauge theory on the Minkowski boundary is obtained as the dual to the AdS$_5$ space-time with a black hole embedded in the bulk, whose event horizon is located at $\chi_0 = (\pi T)^{-1}$. Since the horizon extends to infinity in the Minkowski space-time variables $({\bf x},t)$, one often speaks of a ``black brane''. The thermal excitations of the boundary gauge theory can be considered as the holographic image of the Hawking radiation generated by the black hole.

\begin{figure}
\label{fig:frag}
\centerline{\includegraphics[width=0.98\textwidth]{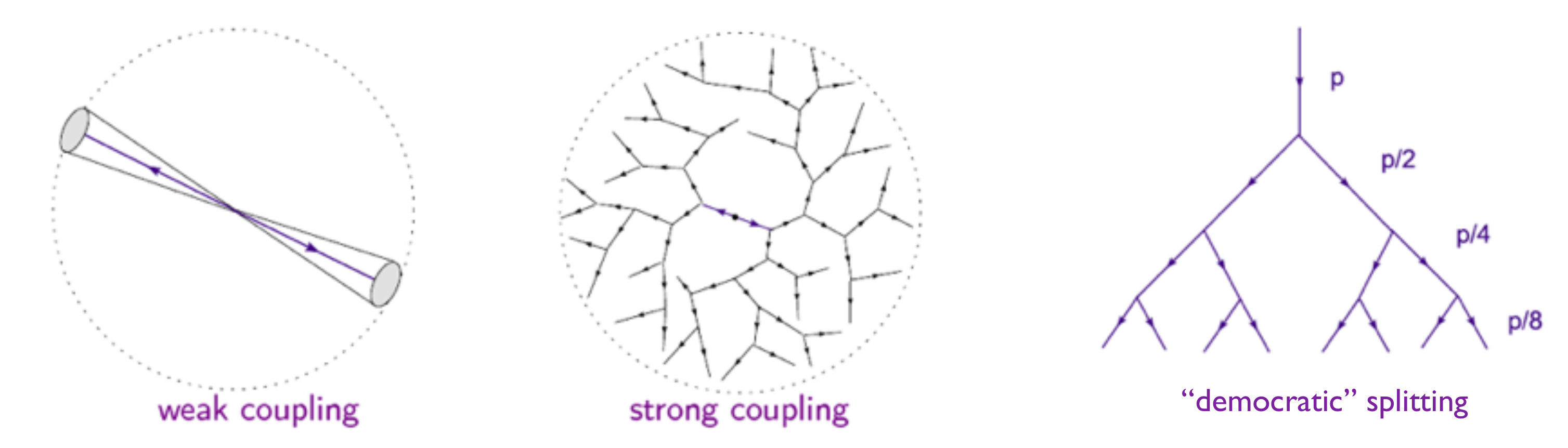}}
\caption{Schematic diagrams showing the dijet-like decay of a localized excitation in a weakly coupled gauge theory (left) and the dendritic decay pattern in the strong coupled gauge theory (center), which produces an approximately isotropic final state. The change of the decay pattern can be attributed to the transition from the dominantly collinear splitting in pQCD to a ``democratic'' splitting pattern (right) in the strongly coupled theory, where both daughter partons  share the energy and momentum of their mother approximately equally.}
\end{figure}

Before we discuss jet quenching in the AdS/CFT context, it is important to clarify that, strictly speaking, jets are not a naturally occurring phenomenon in the strong coupled SYM theory, even in the vacuum. It can be shown that in this theory a highly excited, localized state does not decay into dijet-like configurations as in QCD, but into an isotropic spray of increasingly long-wavelength excitations \cite{Polchinski:2002jw,Hatta:2007cs,Hofman:2008ar}. This can be visualized as the transition from the jet-like fragmentation pattern (left diagram in Fig.~\ref{fig:frag}) in the weakly coupled gauge theory to a dendritic pattern (central diagram in Fig.~\ref{fig:frag}) at strong coupling, which produces a spherical final state in the rest frame of the excitation. The cause of this transition in the kinematic and geometric structure of the final state can be attributed to a change in the fragmentation pattern of hard partons, which is dominated by collinear splittings at weak coupling, but takes on the form of a ``democratic'' splitting cascade (right diagram in Fig.~\ref{fig:frag}), where each daughter parton inherits, on average, half the energy and momentum of its parent.

An excited state with energy $E$ much higher than its virtuality $Q$ will produce a strongly boosted final state, which is isotropic in its own rest frame, but cone-shaped in the lab frame. However, its cone profile will look radically different from that of a pQCD jet: Most of the produced quanta are located near the periphery of the cone, at angles of order $\theta \sim Q/E$, rather than near the core of the cone. It is also important to appreciate that such a ``jet'' would not be naturally produced in the analogue of a heavy ion collision at strong coupling, because the transverse momentum of any localized excitation is very small. In pQCD, every hard jet is accompanied by its recoil jet, which means that the original excitation producing the dijet event has virtuality $Q \approx E$.

In order to discuss jet quenching in the strongly coupled gauge theory, we thus have to assume that strong coupling does not apply to the original process of hard scattering. This makes sense, because we are really interested in exploring the infrared, strongly coupled aspects of QCD, and the QCD coupling is small at short distances. We will simply assume that strong coupling dynamics only sets in after an energetic parton moving through a hot QCD plasma has reduced its virtuality $Q$ down to a modest value $Q \ll E$, of the order of the thermal scale $T$ or the saturation scale $Q_s$ of the medium. At present, we do not have a rigorous theoretical framework to describe this scenario; we will simply assume it.

\section{Energy loss of a fast parton}
\label{sec:eloss}

\begin{figure}
\centering
\begin{minipage}[b]{0.45\textwidth}
\includegraphics[width=0.9\textwidth]{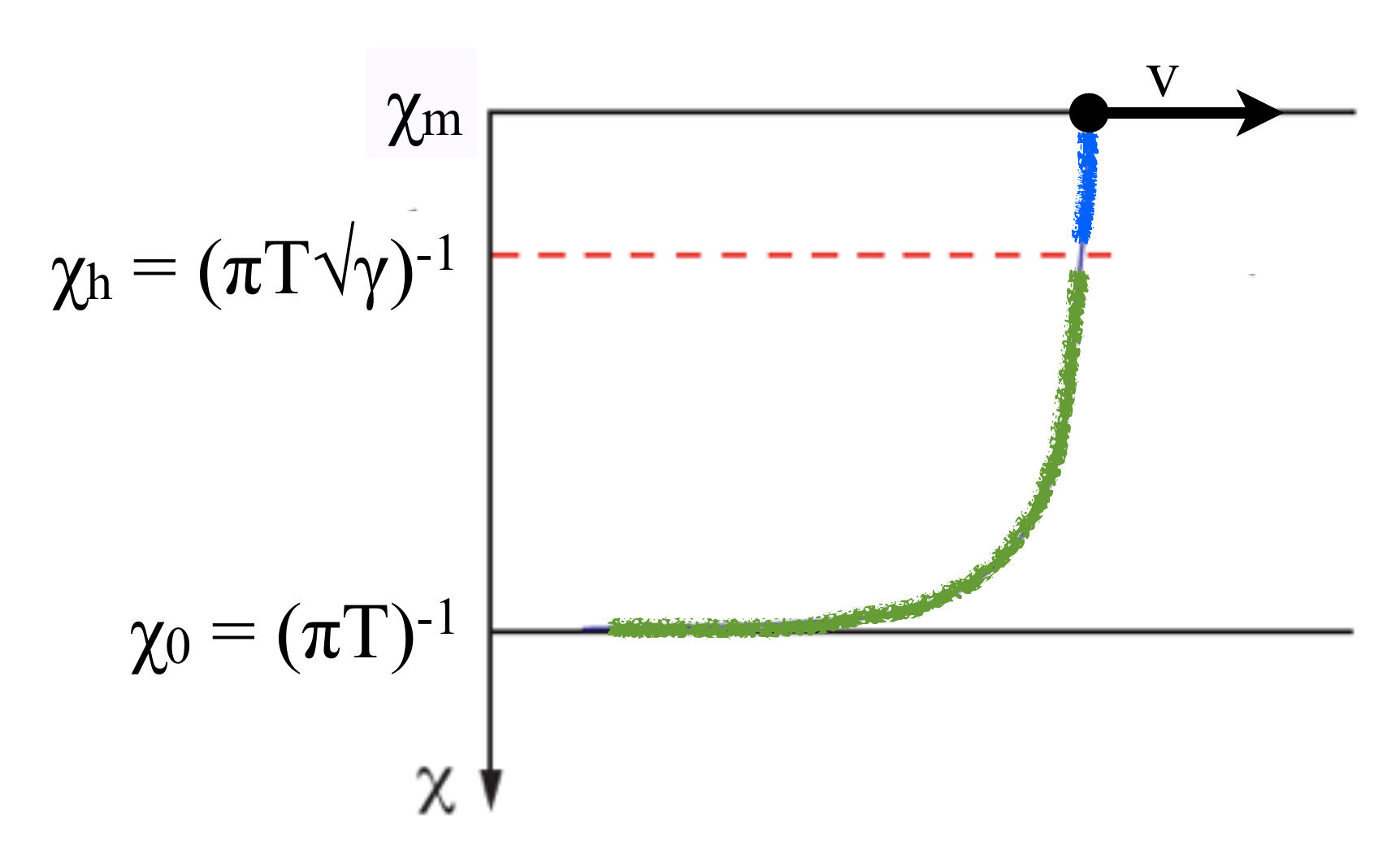}
\caption{Schematic diagram showing the trailing string that dangles off a moving heavy quark into the AdS bulk and asymptotically merges into the black horizon. The string is anchored to a D7-brane located at $\chi_m = \sqrt{\lambda}/m$ near the Minkowski boundary. The part of the string above $\chi_h = (\sqrt{\gamma}T)^{-1}$ represents the comoving color field of the quark; the part below $\chi_h$ is causally disconnected from the quark and represents the radiation field \cite{Dominguez:2008vd}.}
\label{fig:string}
\end{minipage}
\hspace{0.09\textwidth}
\begin{minipage}[b]{0.45\textwidth}
\includegraphics[width=0.95\textwidth]{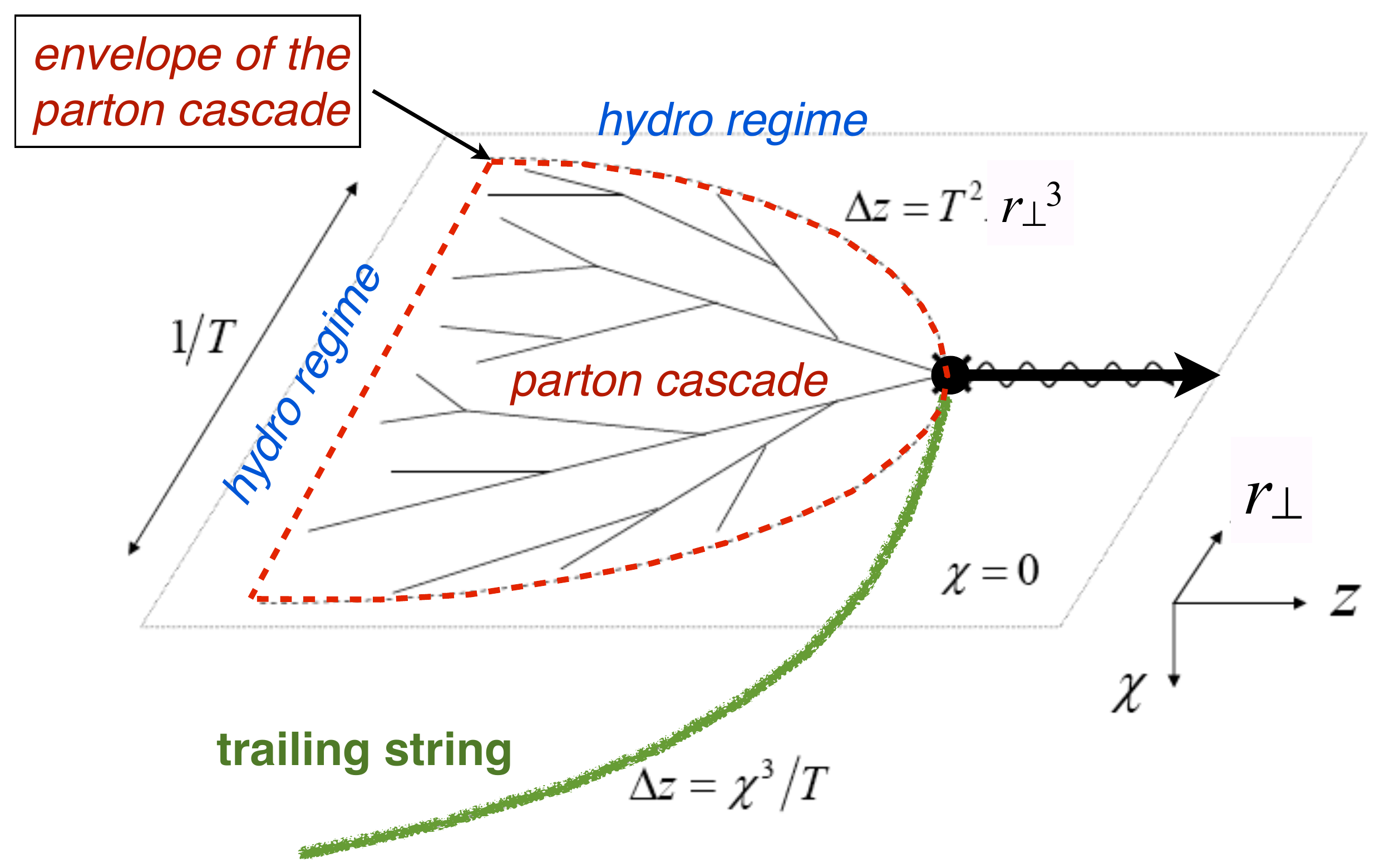}
\caption{The holographic image of the trailing string is a parton cascade in Minkowski space describing the fragmentation of the field energy radiated from the moving quark. The cascade reaches thermal energies at a distance of order $1/T$; beyond this scale the dynamics of energy loss is described by viscous hydrodynamics.}
\label{fig:cascade}
\end{minipage}
\end{figure}

The simplest configuration for the purpose of studying energy loss in the AdS/CFT theory is that of a heavy quark with mass $m \gg T$ moving through the gauge plasma at constant speed $v$. This scenario, in which the heavy quark is represented as the endpoint of a string in the bulk, which is anchored to a D7-brane near the Minkowski boundary, a has been studied by many authors, see e.g., \cite{Gubser:2006qh,Herzog:2006gh,Gubser:2007xz,Chesler:2007an,Chesler:2007sv}. The string represents the holographic image of the color field surrounding the moving quark. As the speed of the quark is increased, the string more and more trails the moving quark as it dangles into the depth of the AdS$-5$ space until it asymptotically merges with the black hole horizon, as shown in Fig.~\ref{fig:string}. 

\setcounter{figure}{4}

The speed of propagation of perturbations along the string is limited; as a consequence, only the upper part of the string above $\chi_h = (\sqrt{\gamma}T)^{-1}$ is causally connected to the moving quark \cite{Dominguez:2008vd}. This part of the string represents the comoving color field of the quark; the part below $\chi_h$ is causally disconnected from the quark and represents the radiation field. Recalling that $1/\chi$ is a measure of the virtuality of field excitations, it makes sense to identify $1/\chi_h$ with the saturation scale of the medium as seen by the moving quark. Field modes with virtuality $Q<Q_s$ are radiated away due to interactions with the medium; those with larger virtuality remain virtual.  The string can thus be interpreted as the holographic image of a partonic fragmentation process describing the cascade of field energy liberated from the quark into modes with ever increasing wavelength, as illustrated in Fig.~\ref{fig:cascade}, starting from the saturation scale $Q_s$ and ending at the thermal scale $1/T$.  One also finds that the rate of energy loss of the heavy quark is $dE/dt \sim - \sqrt{\lambda} Q_s^2$, which has the same form as in pQCD, with $\sqrt{\lambda}$ taking the role of the QCD coupling $\alpha_s N_c$.

\begin{figure}
\centerline{\includegraphics[width=0.4\textwidth]{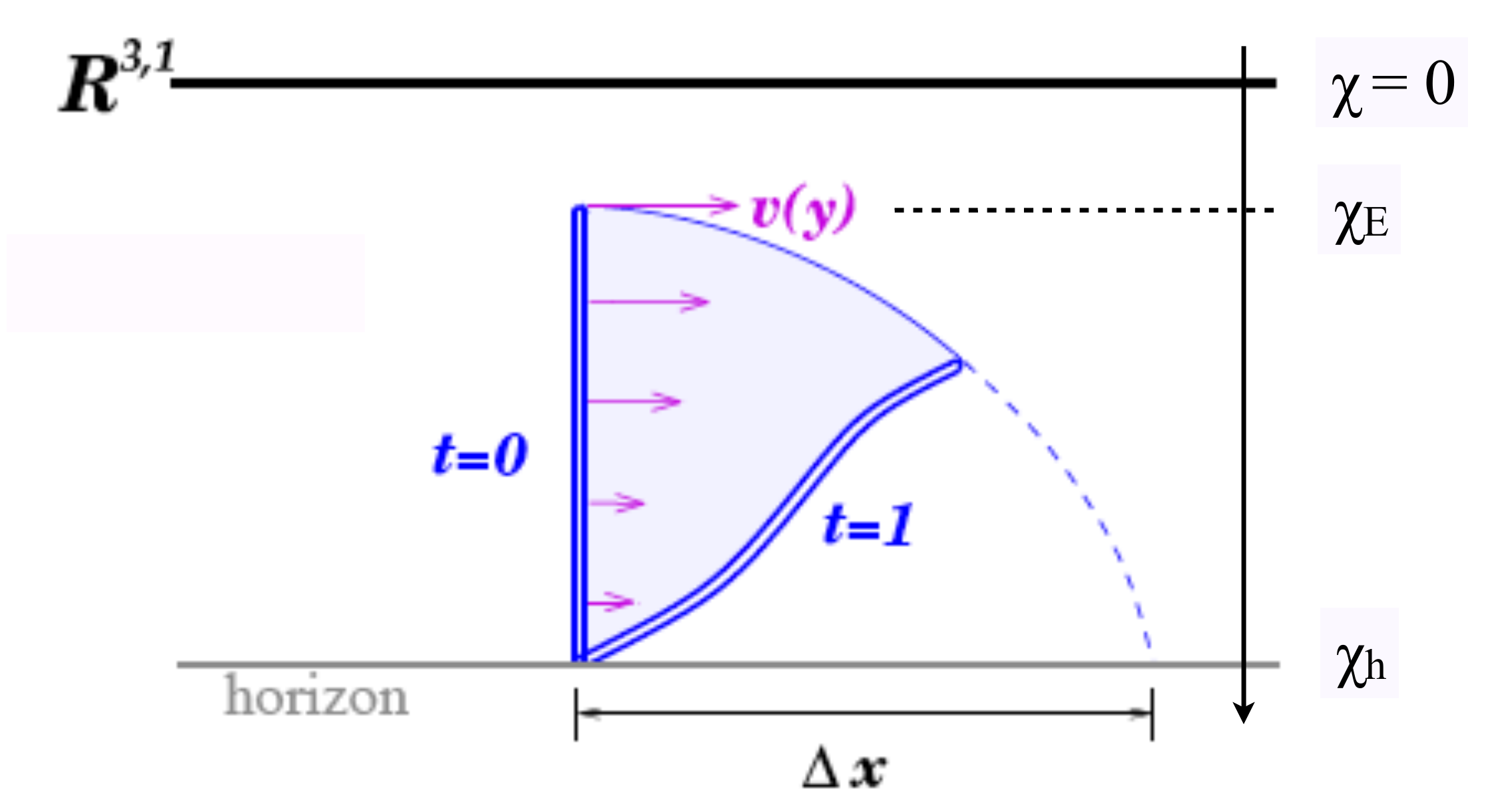}}
\caption{An energetic gluon is modeled in AdS space as a doubled string that emerges from the black horizon. Due to the infinite red shift, the string remains ``anchored'' to a point at the horizon while its turning point at $\chi_E \sim 1/E$ moves rapidly. The string falls toward the event horizon in a rotating motion. Its disappearance into the black hole is the holographic analogue of thermalization in the dual gauge theory.}
\label{fig:double}
\end{figure}

The energy loss of light partons, especially gluons, can be modeled in the following way \cite{Gubser:2008as}: An energetic gluon excitation of energy $E \gg T$ can be represented by a ``doubled up'' string emerging upwards from the black hole horizon, rising up to $\chi_E \sim 1/E$ and then returning to the horizon, as shown in Fig.~\ref{fig:double}. Due to the infinite red shift at the event horizon, the string remains ``anchored'' to a point on the horizon while its turning point at $\chi_E \sim 1/E$ moves rapidly forward. The string thus falls toward the event horizon in a rotating motion and disappears within a finite distance. Its disappearance into the black hole is the holographic analogue of thermalization in the dual gauge theory. The maximal distance covered by the string corresponds to the stopping distance of the energy carried by the hard gluon. A similar calculation can be performed for an energetic supergravity configuration carrying baryon number and thus representing a light quark \cite{Chesler:2008uy}. In both cases the maximal stopping distance given by \cite{Gubser:2008as,Chesler:2008uy}
\be
\ell_{\rm stop}(E) = \frac{\cal C}{T}  \left(\frac{E}{T\sqrt{\lambda}}\right)^{1/3} ,
\ee
with ${\cal C}_g \approx 1/\pi$ for gluons and ${\cal C}_q \approx 1/2$ for light quarks. This result should be compared with the pQCD formula for the stopping distance \cite{Arnold:2009ik}
\be
\ell_{\rm stop}^{\rm (pQCD)}(E) = \frac{\cal C}{\alpha_s^2T}  \left(\frac{E}{TL}\right)^{1/2} ,
\ee 
where $L$ is an energy dependent logarithm. For large energies $E$, the stopping distance in the strongly coupled theory is parametrically shorter. This reflects a more rapid growth of the medium induced virtuality of the parton with time, which we will encounter again when we inspect the length dependence of the energy loss.

As the gluon string approaches the event horizon, its energy loss rate accelerates (the rotating falling motion becomes more and more vertical) leading to a spike in the energy deposition reminiscent of the Bragg peak \cite{Chesler:2008uy}. The same phenomenon occurs in the pQCD scenario because, in addition to the primary parton, also the radiated gluons contribute to the energy deposition into the thermal medium in a ``crescendo'' like manner \cite{Neufeld:2009ep,Qin:2009uh}.

\section{Energy loss phenomenology}
\label{sec:pheno}

\begin{figure}[h]
\centerline{\includegraphics[width=0.4\textwidth]{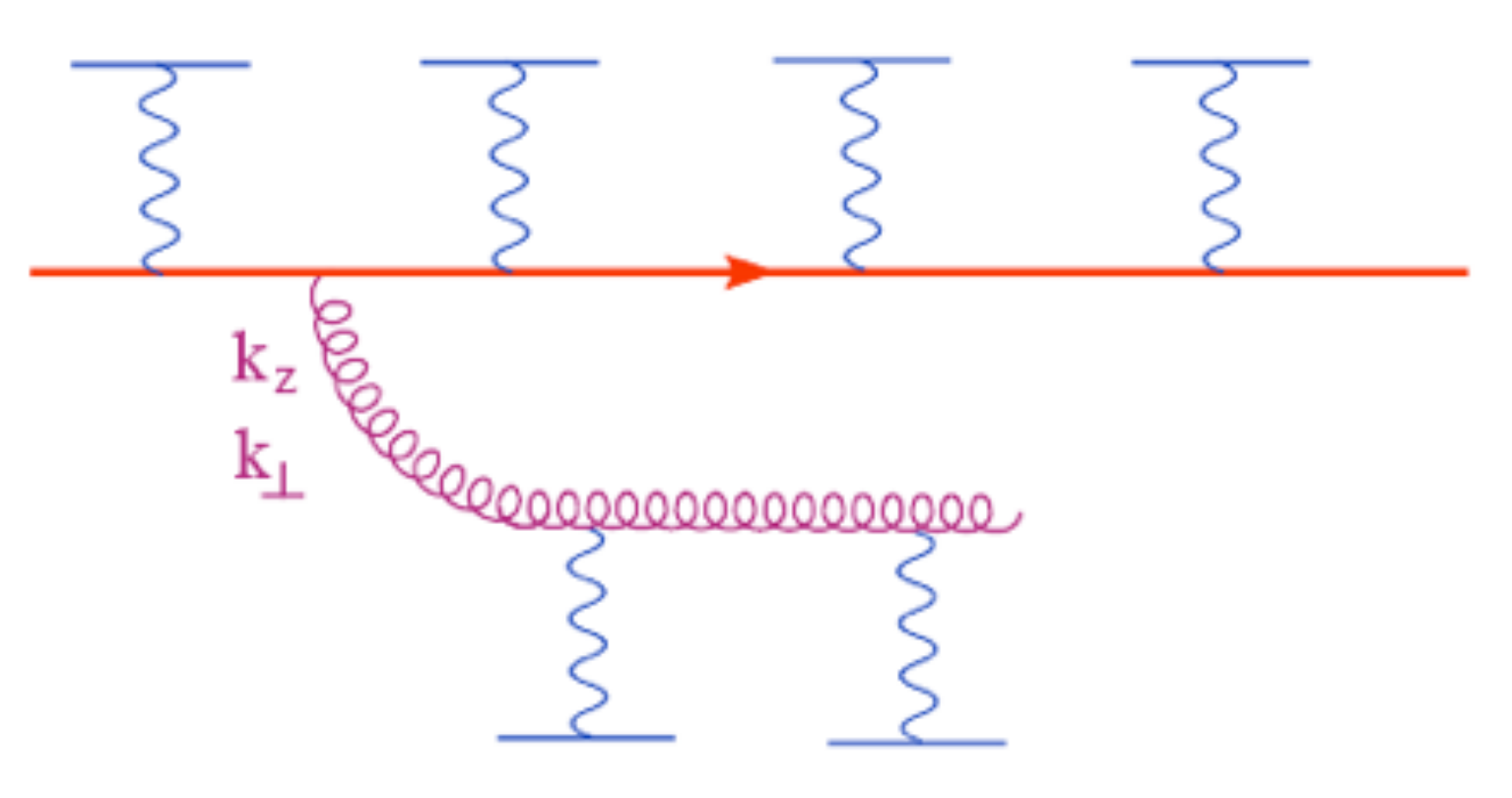}
\hspace{0.1\textwidth}
\includegraphics[width=0.4\textwidth]{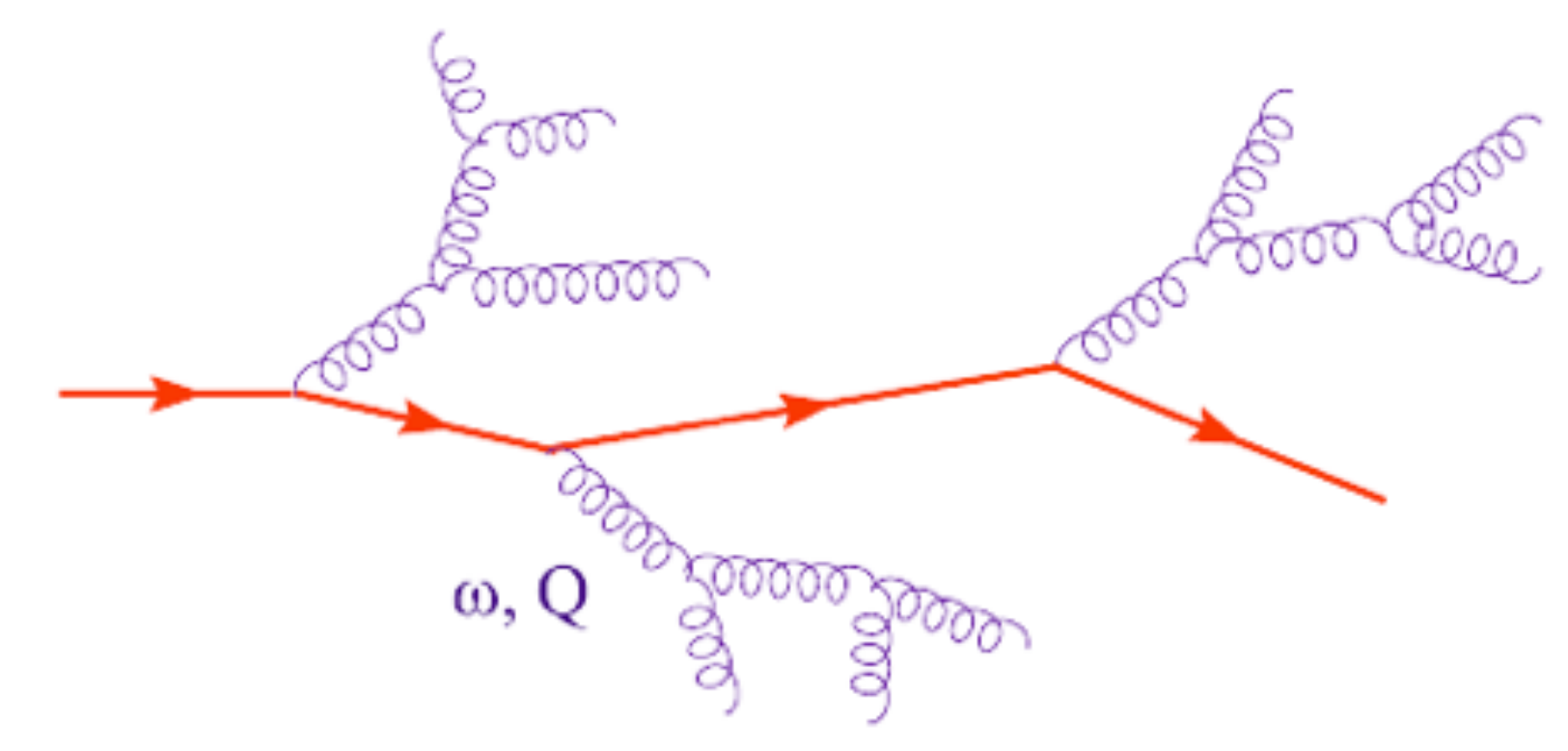} }
\caption{Branching and scattering processes involved in the virtuality evolution of, and radiation by, a parton at weak and strong coupling. At weak coupling gluons are liberated from the hard parton by multiple scattering in the medium (left graph). In the strongly coupled theory, gauge field quanta are liberated by ``democratic'' splittings in a multiple branching cascade (right graph).}
\label{fig:rad}
\end{figure}

Let us recapitulate the differences between parton energy loss in pQCD and in the strongly coupled SYM theory (see Fig.~\ref{fig:rad}):
\begin{itemize}
\setlength{\itemsep}{0pt}
\item In pQCD gluons are liberated from the hard parton by multiple scattering in the medium. The radiation rate is governed by the rate of transverse momentum diffusion of the gluons. The virtuality squared of a parton grows linearly with the distance $L$ covered: $Q_s^2 \sim \hat{q}L \sim \alpha_s T^3 L$.
\item In the strongly coupled SYM theory, gauge field quanta are liberated by ``democratic'' splittings in a multiple branching cascade. The partons are kept off-shell by the thermal force of magnitude $T^2$ exerted by the medium on the partons. Since the work done always has the same sign, the virtuality decreases linearly with the distance: $dQ/dt \sim T^2$ \cite{Hatta:2008tx,Dominguez:2008vd,Albacete:2008ze}.
\end{itemize}

The virtuality evolution $Q(t)$ of a jet differs between pQCD and strongly coupled SYM (AdS/CFT) both in the vacuum and in the medium. The same is true for the parton splitting functions $P(z)$, which are characterized by collinear singularities in pQCD, but reflect a ``democratic'' distribution of energy and momentum in AdS/CFT. The virtuality evolution with time of a hard parton is schematically shown in Table~\ref{tab:1}.
\begin{table}[h]
\centering
\begin{tabular}{|c|c|c|}
\hline
~~ & Vacuum & Medium \cr
\hline
~~ & ~~ & ~~ \cr
pQCD & $Q(t)^2 \sim \frac{E\hbar}{2t}$ & $Q(t)^2 \sim \hat{q} t \sim T^3 t$ \cr
~~ & ~~ & ~~ \cr
AdS/CFT & $Q(t)^2 \sim \left(\frac{E\hbar}{Q_0t}\right)^2$ & $dQ/dt \sim -T^2$ \cr
~~ & ~~ & ~~ \cr
\hline
\end{tabular}
\caption{Virtuality evolution of a parton in vacuum and in medium for the weakly coupled pQCD and the strongly coupled AdS/CFT scenarios.}
\label{tab:1}
\end{table}

\begin{figure}[h]
\centerline{\includegraphics[width=0.4\textwidth]{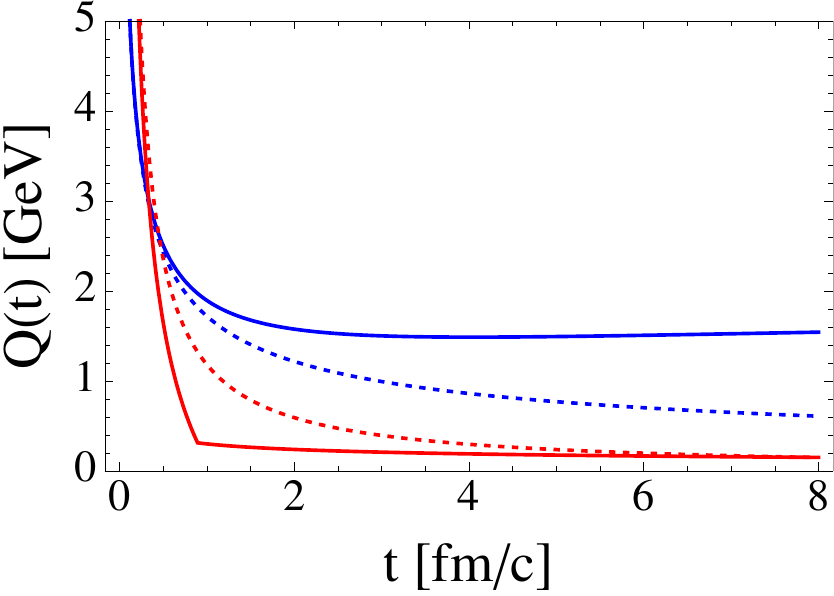}
\hspace{0.1\textwidth}
\includegraphics[width=0.4\textwidth]{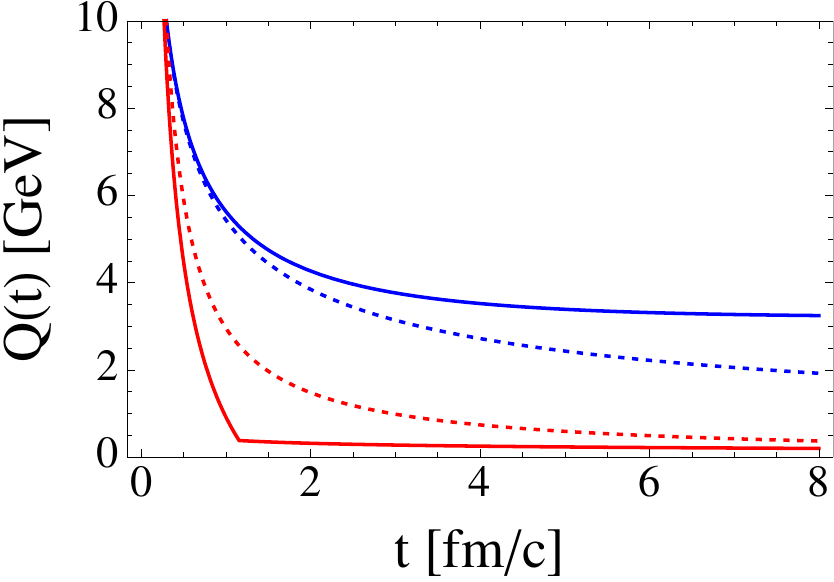} }
\caption{The left panel shows the virtuality evolution of a gluon with energy $E=30$ GeV and initial virtuality $Q_0=5$ GeV in a medium with temperature $T_0=300$ MeV at time $t_0=1$ fm/c, typical for the quark-gluon plasma formed at RHIC (red lines for AdS/CFT; blue lines for pQCD). The dotted lines show the vacuum evolution. The right panel shows the virtuality evolution of a 300 GeV gluon with an initial virtuality  $Q_0=20$ GeV for $T_0(t_0)=400$ MeV, expected to be typical for LHC energies. Same color scheme as in the left panel.}
\label{fig:virt}
\end{figure}

As an example, let us consider the virtuality evolution of a gluon with initial energy $E=30$ GeV, initial virtuality $Q_0=5$ GeV in a medium with temperature $T_0=300$ MeV at time $t_0=1$ fm/c typical for the quark-gluon plasma formed at RHIC. We assume that the medium expands longitudinally in a boost invariant manner, so that $T(t) = T_0(t_0/t)^{1/3}$. We use the standard pQCD expression for $\hat{q}$ as given by \cite{Arnold:2009ik}.  We also assume for simplicity that the vacuum and medium contributions to $Q^2$ are additive. The virtuality evolution is shown in the left panel of Fig.~\ref{fig:virt} by solid lines (red for AdS/CFT; blue for pQCD). The vacuum evolution of the virtuality is shown by the dotted lines. One notices two main features. One is that the in-medium contribution to the parton virtuality begins to dominate after about 0.5 fm/c in the AdS/CFT scenario, but significantly later (for $t>2.5$ fm/c) in the pQCD scenario; in addition, the modifications are opposite in sign. Because gluon radiation in the pQCD scenario is rare, the medium tends to increase the parton virtuality; in the AdS/CFT scenario, the medium rapidly reduces the virtuality by splitting \cite{Iancu:2008sp}. The weakly and strongly coupled theories differ drasctically in their predictions for $t>2$ fm/c. 

The right panel of Fig.~\ref{fig:virt} shows the virtuality evolution of a 300 GeV gluon with an initial virtuality  $Q_0=20$ GeV for $T_0(t_0)=400$ MeV, which may be characteristic for heavy-ion collisions at the LHC. Here one notes two things. Firstly, the gluon virtuality in pQCD exceeds that in AdS/CFT until $t\approx 3$ fm/c, and secondly, it takes almost 5 fm/c for the vacuum contribution to the parton virtuality to fall below the in-medium contribution in the pQCD scenario. This effect is due to the collinear splitting in pQCD, which reduces the parton energy only gradually and thus leads to an increase in time dilation as the virtuality drops. This means that the very energetic parton hardly notices the medium for the first $3-4$ fm of its path length. On the other hand, in the AdS/CFT scenario, parton energy and virtuality decrease proportionally implying a fixed time dilation effect. 

The difference between the virtuality evolution within the medium in pQCD and strongly coupled AdS/CFT lies at the root of the different length dependence of the leading parton energy loss rate ($dE/dx \sim L$ versus $dE/dx \sim L^2$).  A phenomenological study of the effects of the different length dependence was recently performed by Marquet and Renk \cite{Marquet:2009eq}. These authors found that the quadratic length dependence of the AdS/CFT scenario gives a better representation of the emission angle dependence of the suppression factor $R_{AA}(\pi^0)$ with respect to the reaction plane in semi-central Au+Au collisions. On the other hand, the momentum dependence of the back-to-back suppression factor $I_{AA}$ measured by STAR slightly favors the pQCD scenario. The data thus do not allow for a firm conclusion at this time on the basis of leading hadron suppression effects alone.

Let us next discuss the splitting functions. In pQCD the gluon-gluon splitting function is
\be
P_{g\to gg}^{\rm (pQCD)} = 2N_c\frac{(1-z(1-z))^2}{z(1-z)} .
\ee
The gluon splitting function in the strongly coupled SYM theory is not known analytically, but it is clear that its most probable value must be $z=1/2$, implying a ``democratic'' splitting pattern \cite{Hatta:2008tx}. The splitting function can, in principle, be derived from the analysis of the energy flow on the Minkowski space boundary. In lack of an exact result, we use the ansatz
\be
P_{g\to gg}^{\rm (AdS/CFT)} = \frac{42}{5}\,N_c = {\rm const.} ,
\ee
where the numerical factor is chosen such that $\int_0^1 dz\, P(z)\, z(1-z)$ is the same for both splitting functions. 

In order to exhibit the differences between the two expressions with respect to the parton fragmentation cascade, we calculate the total integrated splitting probability $\Delta(z,t)$ as a function of time. We incorporate the LPM effect by demanding that the splitting time $t$ be larger than the formation time $t_f = z(1-z)E/k_T^2$, and we require the transverse momentum constraint $k_T \leq z(1-z)E$. The integrated splitting probability is defined as
\be
\Delta(z,t) = \frac{\cal C}{2\pi} \int \frac{dk_T^2}{k_T^2}\, P(z)\, \theta\big(z(1-z)E-k_T\big)\, 
  \theta(k_T - k_{T,{\rm min}})\,
  \theta\left(t-\frac{z(1-z)E}{k_T^2}\right) .
\label{eq:Delta}
\ee
Formally, $\Delta$ can be regarded as the kernel of the logarithm of the Sudakov factor, $S(t) = \exp\big(-\int_0^{1/2} \Delta(z,t)\, dz\big)$ which denotes the no-splitting probability up to time $t$. Its time derivative is the kernel of the time evolution equation for the fragmentation function:
\be
\frac{dD(y,t)}{dt} = \int_0^1 \frac{dz}{z}\, \frac{d\Delta(z,t)}{dt}\, D\left(\frac{y}{z}\right) ,
\ee
which can be considered as the DGLAP equation for the fragmentation function expressed in terms of time $t$ instead of virtuality $Q^2$. The pQCD value for the prefactor in (\ref{eq:Delta}) is ${\cal C}_{\rm pQCD} = \alpha_s$; the correct value for the AdS/CFT case is unknown, but it is plausibly proportional to $\sqrt{\lambda}$. To be conservative, we here choose ${\cal C}_{\rm AdS/CFT} = 1$. We also set $\alpha_s=0.3$ and $k_{T,{\rm min}}=1$ GeV.

\begin{figure}
\centering
\begin{minipage}[b]{0.49\textwidth}
\includegraphics[width=0.83\textwidth]{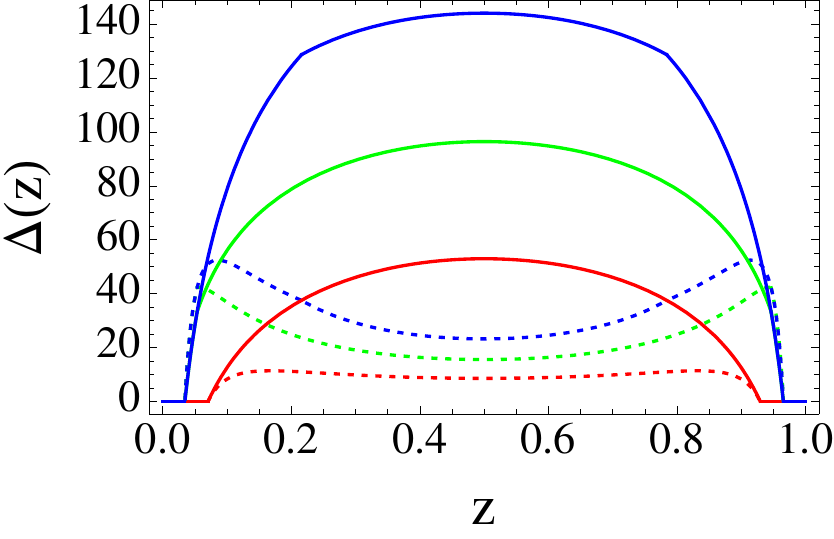}
\caption{The integrated splitting function $\Delta(z,t)$ for a gluon of fixed energy $E=30$ GeV and initial virtuality $Q_0=5$ GeV. The dashed lines show the pQCD result; the solid lines show the results for the AdS/CFT splitting scheme. From bottom to top, the red, green, and blue lines represent the results for $t=0.1,0.3,1.0$ fm/c respectively. The difference between the pQCD and AdS/CFT splitting patterns are striking.}
\label{fig:split1}
\end{minipage}
\hspace{0.05\textwidth}
\begin{minipage}[b]{0.45\textwidth}
\includegraphics[width=0.95\textwidth]{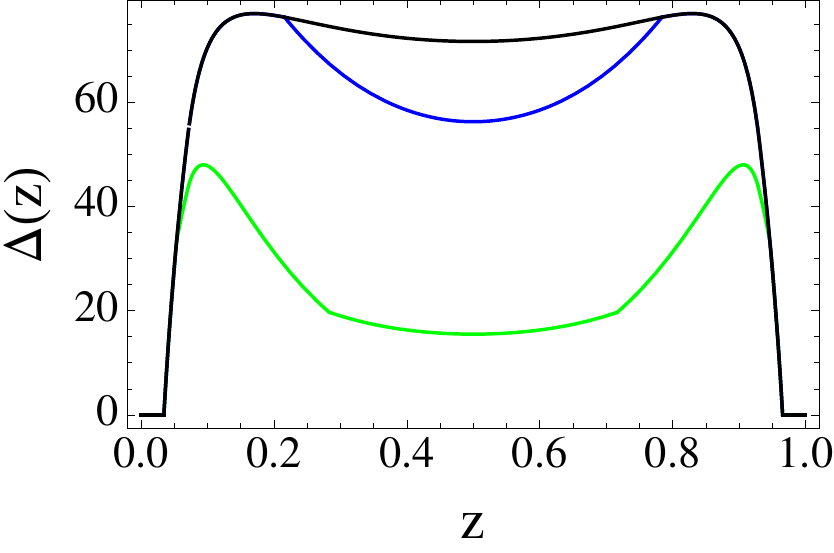}
\caption{The integrated splitting function $\Delta(z,t)$ for $k_c=2$ GeV and times $t=0.3,1,3$ fm/c (from bottom to top shown as green, blue, and black curves), including both the pQCD and the AdS/CFT contributions, for the same gluon energy and initial virtuality as in Fig.~\ref{fig:split1}.}
\label{fig:split2}
\end{minipage}
\end{figure}

Figure~\ref{fig:split1} shows $\Delta(z,t)$ for a gluon of fixed energy $E=30$ GeV and initial virtuality $Q_0=5$ GeV. The dashed lines show the pQCD result; the solid lines show the results for the AdS/CFT splitting scheme. From bottom to top, the red, green, and blue lines represent the results for $t=0.1,0.3,1.0$ fm/c respectively. The difference between the pQCD and AdS/CFT splitting patterns are striking. The AdS/CFT splitting pattern will result in a very different fragmentation function and lead to a much faster degradation of the energy of the leading parton than the collinear pQCD splitting pattern.

We must remember, however, that we do not expect the AdS/CFT (strong coupling) pattern to set in immediately, but only after the virtuality of the parton has evolved down into the strong coupling regime. As Fig.~\ref{fig:virt} shows, this takes some time. A more realistic result would thus be obtained by using a virtuality dependent splitting function, such as
\be
P(z,k_T) = P_{g\to gg}^{\rm (pQCD)}(z)\, \theta(k_T-k_c) + P_{g\to gg}^{\rm (AdS/CFT)}(z)\, \theta(k_c-k_T) ,
\ee
where $k_c$ denotes the virtuality threshold between the two regimes. For illustration, we show in Fig.~\ref{fig:split2} the result for the integrated splitting function $\Delta(z,t)$ for $k_c=2$ GeV and times $t=0.3,1,3$ fm/c (from bottom to top shown as green, blue, and black curves) including both the pQCD and the AdS/CFT contributions. The strong coupling scheme only operates in the virtuality range $1~{\rm GeV} \leq Q \leq 2$ GeV; this reduces the extent of the modification of the fragmentation cascade. 

Note that the effect of the medium on the time evolution of the parton virtuality is not included in Figs.~\ref{fig:split1} and~\ref{fig:split2}. The medium prolongs the time spent by the partons in the virtuality range where strong coupling applies. One can thus expect that the medium will enhance the difference between a pQCD and a hybrid pQCD--AdS/CFT fragmentation function. It would be interesting to perform such a simulation.

A somewhat different hybrid approach to jet quenching was proposed by Liu {\em et al.} \cite{Liu:2006he,Liu:2006ug} (LRW, see also \cite{DEramo:2010ak,DEramo:2010xk}). These authors adopt the pQCD definition of the transverse momentum diffusion parameter $\hat{q}$, but evaluate it nonperturbatively in the strongly coupled $N=4$ SYM theory. The LRW approach may provide an explanation for the unexpectedly large value of $\hat{q}$ required to describe the RHIC data; it may also explain the absence of a logarithmic energy dependence of $\hat{q}$ advocated by some \cite{Qin:2009gw}. The LRW approach assumes that strong coupling only operates at thermal scales relevant to the medium, but never at the virtuality scales reached by the jet. In view of the virtuality evolution shown in the left panel of Fig.~\ref{fig:virt}, where $Q(t)$ always exceeds 2 GeV, this assumption may, indeed, be warranted as long as one considered only the fate of the leading parton.

Finally, it is worth mentioning the hadronic energy loss mechanism recently proposed in \cite{CasalderreySolana:2009ch,CasalderreySolana:2010xh}. It is based on the insight that heavy quark bound states, e.~g.\ the $J/\psi$, which may exist above $T_c$ in a strongly coupled medium, will propagate with a limited speed. This is expected to be a general effect in any gauge theory with a gravity dual, because it is caused by the gravitational red shift effect of the heavy bound state in the AdS/CFT dual description. A fast light parton, which travels at near the speed of light, will then be able to emit such heavy mesons as Cherenkov radiation. For details, the reader is referred to Casalderrey-Solana's talk \cite{CasalderreySolana:2010HP}.

\section{Summary}
\label{sec:summary}

In summarizing, it is useful to recapitulate the salient questions about strong coupling in jet quenching and what we know about their answers:
\begin{itemize}
\setlength{\itemsep}{0pt}
\item Are jets formed in the presence of a dense QGP medium?\\
{\em Answer:}  Yes. This implies that the QCD coupling is weak before the virtuality reaches medium scale. (This is a natural prediction in view of the asymptotic freedom of QCD.)
\item Is the quark-gluon plasma strongly coupled?\\
{\em Answer:}  Probably, Yes. The RHIC data indicate that jet quenching is stronger than expected from a
perturbative quark-gluon plasma. (This is also the conclusion reached from the small value of the shear viscosity demanded by the data.) Note that this is not a problem for pQCD jet quenching theory, because it does not require the medium structure to be perturbative.
\item  Is the parent parton strongly coupled to the medium after reaching the medium virtuality?\\
{\em Answer:}  We do not know. 
\item Are the radiated gluons strongly coupled?\\
{\em Answer:}  We do not know. 
\end{itemize}

This brings us to the question: What is needed to make progress? Here are some things that could be done by careful comparisons between high statistics data and high quality simulations of jet quenching using either the PQCD or the AdS/CFT paradigm, or hybrid approaches:
\begin{itemize}
\setlength{\itemsep}{0pt}
\item Establish the length dependence of leading energy loss.
\item Establish the energy dependence of leading parton energy loss.
\item Measure the medium modification of the angular (jet shape) and momentum distribution (fragmentation function) of the radiated energy.
\item Measure the quark mass dependence of leading parton energy loss.
\end{itemize}
There exist many opportunities to pursue these investigations at RHIC and LHC. Specific tools at hand include: jets tagged by energy or flavor; the angle dependence of jet quenching with respect to the reaction plane; the system size dependence of jet quenching; the analysis of the longitudinal and transverse structure of fully reconstructed jets. Finally, it seems clear that one or two advanced detectors with world-class jet measurement capabilities (acceptance, resolution, counting rate) will be needed at RHIC in order to realize these opportunities.

\section{Acknowledgments}

This work was supported in part by a grant from the U.~S.~Department of Energy (DE-FG02-05ER41367). I thank the organizers for the opportunity to give this talk, and J.~Casalderrey-Solana, E.~Iancu, Yu.~Kovchegov, A.~Majumder, and D.~Teaney for helpful discussions.



\end{document}